\documentclass{article}

\usepackage{microtype}
\usepackage{graphicx}
\usepackage{subfigure}
\usepackage{booktabs} % for professional tables
\usepackage{setspace}
\usepackage{hyperref}

\usepackage[accepted]{icml2023}
\usepackage{amsmath}
\usepackage{amssymb}
\usepackage{mathtools}
\usepackage{amsthm}

\usepackage[capitalize,noabbrev]{cleveref}

\theoremstyle{plain}

\theoremstyle{definition}

\theoremstyle{remark}

\usepackage{todonotes}

\icmltitlerunning{Diffusion Models for Probabilistic Deconvolution of Galaxy Images}

\begin{document}

\twocolumn[
\icmltitle{Diffusion Models for Probabilistic Deconvolution of Galaxy Images}



\icmlsetsymbol{equal}{*}

\begin{icmlauthorlist}
\icmlauthor{Zhiwei Xue}{equal,umich}
\icmlauthor{Yuhang Li}{equal,umich}
\icmlauthor{Yash Patel}{umich}
\icmlauthor{Jeffrey Regier}{umich}
\end{icmlauthorlist}

\icmlaffiliation{umich}{Department of Statistics, University of Michigan}

\icmlcorrespondingauthor{Jeffrey Regier}{regier@umich.edu}

\icmlkeywords{Machine Learning, ICML}

\vskip 0.3in
]



\printAffiliationsAndNotice{\icmlEqualContribution} % otherwise use the standard text.

\begin{abstract}
Telescopes capture images with a particular point spread function (PSF). Inferring what an image would have looked like with a much sharper PSF, a problem known as PSF deconvolution, is ill-posed because PSF convolution is not an invertible transformation.
Deep generative models are appealing for PSF deconvolution because they can infer a posterior distribution over candidate images that, if convolved with the PSF, could have generated the observation. However, classical deep generative models such as VAEs and GANs often provide inadequate sample diversity. As an alternative, we propose a classifier-free conditional diffusion model for PSF deconvolution of galaxy images. We demonstrate that this diffusion model captures a greater diversity of possible deconvolutions compared to a conditional VAE.
\end{abstract}

\section{Introduction}
\label{intro}
High-fidelity galaxy models are important for deblending~\cite{melchior2021challenge}, analyzing lens substructure~\cite{mishra2022strong}, and validating the analysis of optical surveys~\cite{korytov2019cosmodc2}. Traditional galaxy models rely on simple parameteric profiles such as Sersic profiles \cite{sersic1963influence}.
However, these models fail to capture the rich structures that are visible in modern surveys. As a result, there is growing interest in using deep generative models, such as variational autoencoders (VAEs), to represent galaxies~\cite{regier2015deep,castelvecchi2017astronomers,lanusse2021deep}. 

Deep generative models of galaxies are fitted with images that have been observed with a particular point-spread function (PSF). It is thus necessary to account for the PSF in fitting these galaxy models, to disentangle the measurement process from the physical reality.

Because PSF convolution is not an invertible transformation, multiple deconvolved images are compatible with the observed image.
% Reviewer 2: I believe PSF deconvolution alone can be well-posed if the PSF is known exactly and respect certain properties. Deconvolution is not invertible when information is lost, which is often caused by downsampling after blurring or after noise is added to the measurements (which is always present, but impacts astronomical images most significantly). What are the thoughts of the author on this?
Traditional deconvolution methods produce only one deconvolved image that is compatible with the image~\cite{vojtekova2021learning}. Deep generative models are an appealing alternative because they infer a distribution of deconvolved images that are compatible with an observation. Although conditional VAEs and conditional GANs~\cite{schawinski2017generative,fussell2019forging,lanusse2021deep} can provide a distribution of deconvolved images, both are known to produce insufficient diversity in their outputs \cite{salimans2016improved}. 

Diffusion models are a recently developed alternative to VAEs and GANs that excel at producing diverse samples. Diffusion models have been successfully applied to solve inverse problems \cite{kawar2022denoising,remy2023probabilistic, adam2022posterior, song2022solving}.
% Jeff: In a longer version of this paper, we'd want to say more about these references
However, training diffusion models with PSF convolved data to learn a representation of physical reality that is decoupled from the measurement process is not as straightforward as with a VAE. If we simply add a PSF convolution layer to the end of the diffusion model's decoder, as we can with a VAE, training is no longer tractable.
% Reviewer 1: A longer version of the paper could add an explanation why diffusion models can't simply add a PSF convolution after the last layer in the way VAEs can
% Reviewer 2 (a bit off-base here, but has something of a point):  The problem raised about the impossibility of training score-models from corrupted samples in the current paradigm is of high interest and should be emphasized much more in the text. Training a score-model from simulated galaxy images seems like an attractive solution as they form a principled prior (although potentially biased) over galaxy surface brightness profile. What are the authors' thoughts on learning a prior over simulations that do not match exactly the physical reality?

Instead, we propose to model PSF-convolved galaxy images with a classifier-free conditional diffusion model~\cite{ho2021classifier} and to condition on the observed PSF. In training this model, we make use of \textit{paired} data sources, e.g., both ground-based and space-based telescopes. We used a conditional VAE as a baseline and compared the methods using a novel evaluation method. We find that CVAEs tended to produce high percentages of invalid deconvolutions due to missing high-frequency details in reconstructions, resulting in lower sample diversity compared to conditional diffusion models.
Our code is available from \url{https://github.com/yashpatel5400/galgen}

\section{Methods}
\label{method}
Let $x$ denote the observed (PSF-convolved) image, let $y$ denote the latent ``clean'' image, and let $\Pi$ denote the PSF. Then, neglecting pixelation and measurement noise, \mbox{$x = \Pi * y$}. 
We investigate classifier-free conditional diffusion models for solving the deconvolution task and we consider conditional VAEs as a baseline to compare against. 
% Reviewer 2: In the (bayesian) inverse problem literature, 'y' is typically reserved for the measurement and 'x' for the parameters of interest. Inverting them is slightly confusing. I believe it led to some errors in section 2.3, where the authors mention that "no dataset is available with observed y that also contains multiple draws of x for each y". Shouldn't it be the reverse?

\subsection{Conditional VAE}
VAEs model the data distribution as a transformation of a lower-dimensional latent space \cite{kingma2013auto}. An encoder $q_\varphi$ maps the input $x$ to a distribution over a low-dimensional latent expression $z$, which defines an approximate posterior distribution $q_\varphi(z\mid x)$; a decoder $p_\theta$ maps $z$ to the original data space through a generative model $p_\theta(x\mid z)$. 
Conditional VAEs (CVAEs) \cite{sohn2015learning} extend VAEs by conditioning both the encoder and decoder on auxiliary variables $c$, which may be denoted as $q_\varphi(z\mid x, c)$ and $p_\theta(x\mid z, c)$, respectively. 

\begin{figure}
\includegraphics[scale=0.2]{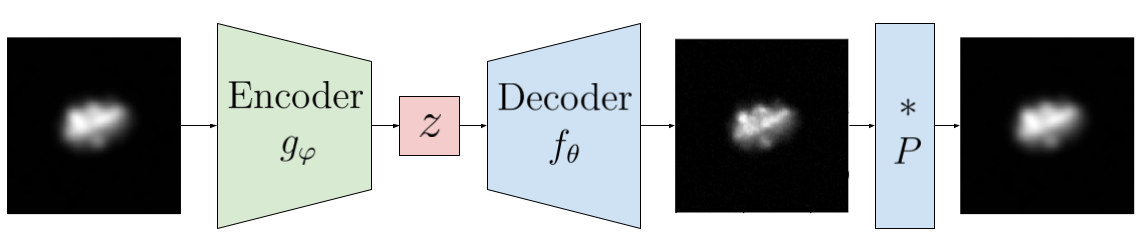}
\centering
\caption{\label{fig:vae_sr} A CVAE is employed with a partially \textit{fixed} decoder consisting of a deterministic convolution with the known PSF.}
\end{figure}

We investigate a CVAE in which $c=\Pi$ and the final layer of the decoder is fixed to be a convolution with the known PSF, as in \citet{lanusse2021deep} and illustrated in Figure \ref{fig:vae_sr}.
With this approach, for each draw from the latent space, a candidate deconvolved image is produced as an intermediate result in the decoder, which is the quantity targeted by inference. For training, we take the loss to be a weighted-variant of the ELBO for the joint distribution $p(x,y\mid z)$:
\begin{equation}
\begin{gathered}
    \mathcal{L}_{\text{CVAE}} :=
    \alpha\ \mathbb{E}_{z\sim q_\varphi(z\mid x)}[\log(p_{\theta}(y \mid z))] \\  
    + \beta\ \mathbb{E}_{z\sim q_\varphi(z\mid x)}[\log(p_{\theta}(x \mid y, z))] \\ 
    - D_{KL}(q_\varphi(z\mid x) || p_{\theta}(z)).
\end{gathered}
\end{equation}
Note that the first term targets the reconstruction of the deconvolved image. The hyperparameter weighting terms $\alpha$ and $\beta$  asymmetrically weight the PSF-convolved and deconvolved reconstructions, a formulation similar to the $\beta$-VAE \cite{higgins2017beta}, which we found to improve the reconstruction of high-frequency details in samples.

\subsection{Conditional Diffusion Models}
Conditional diffusion models, an extension on denoising diffusion probabilistic models \cite{ho2020denoising}, are trained with the following loss function:
\begin{equation}
    \mathcal{L}_{\text{Diff}}
    := \mathbb{E}_{t \sim [1, T], \mathbf{x}_0, \boldsymbol{\epsilon_t}} \left[||\boldsymbol{\epsilon}_t - \boldsymbol{\epsilon}_\theta(\mathbf{x}_t, t,c)||^2 \right],
\end{equation}
where $\boldsymbol{\epsilon}_\theta(\mathbf{x}_t, t,c)$ predicts the noise added to $\mathbf{x}_t$ (the latent variable for time step $t$) and $c$ is the conditioning information. We set $c=(x, \Pi)$.
% Reviewer 2: Insufficient citations are given when describing the denoising score matching loss. Such a training objective arose from the foundational work of Aapo Hyvärinen (2005) and Pascal Vincent (2011). They should thus be cited in addition to Ho et al. (2020). Furthermore, I suggest that the loss be described in more explicit terms. For example: "denoising score-matching training objective" or something along the line of "Fisher divergence between the score model and the Gaussian perturbation kernel".
Therefore, at inference time, to deconvolve the image $x \in \mathbb{R}^{k_1\times k_2}$, a deconvolved image $y$ is sampled by first sampling $x_0\sim\mathcal{N}(0, I_{k_1k_2, k_1k_2})$ and then taking $T$ denoising steps conditioned on $(x, \Pi)$.

% \jeff{This is a pretty minimalist description of the proposed method. In fact, it's more like a subsection of a background section than a proposed method. Is there any more that can be said about it?}

\subsection{Evaluation Metrics}
Recent works such as \citet{hackstein2023evaluation} have investigated metrics for the related task of generating galaxy images. However, the task of galaxy generation is distinct from ours, as we are seeking to produce diverse candidates conditional on a \textit{single} observed image. Thus, simply measuring the recovery of the marginal distribution $p(y)$ of deconvolved images is insufficient to assess performance for our task. Furthermore, no paired reference data is available that provides observations of $x$ paired with multiple draws of $y$ for each $x$.

Instead, to assess the diversity of samples from the posterior $p(y\mid x, \Pi)$, we propose the following metric, where a given $q$ must satisfy the specified constraint:
\begin{equation}\label{eqn:obj}
    \begin{aligned}
    % \max_{q} \quad & 
    & \mathbb{E}_{x\sim p(x)}\left[\mathbb{V}_{y\sim q(y\mid x, \Pi)}[y]\right] \\
    \textrm{s.t.} \quad & \mathbb{E}_{x\sim p(x)}\left[
    \mathbb{E}_{y\sim q(y\mid x, \Pi)}\left[||\Pi * y - x||_2^2\right]\right] < \epsilon.
    % d_F(p(y), q(y)) < \epsilon,
    \end{aligned}
    \end{equation}
Here, $\epsilon$ represents an allowed slack and $\mathbb{V}$ denotes the total variance of $y$ given both $x$ and $\Pi$. By adding the constraint, we ensure that high-scoring methods produce valid deconvolutions.
% \jeff{It'd be good to say something here about why the proposed optimization problem is intuitively a reasonable thing to do.}
% , i.e., the determinant of its covariance matrix. 
% and $d_F$ the Fréchet inception distance (FID). 
To avoid favoring methods that generate images with imperceptible pixel-level variance,
we compute $\mathbb{V}$ over image featurizations, defined by mapping the domain of images $\mathcal{Y}$ to image features $\mathcal{F}$ with a pre-trained InceptionV3 network;
% \jeff{What is ``image featurization''? (a reader probably won't know)}
this idea is inspired by the Fréchet inception distance (FID). That is, for distributions $p(y\mid x)$ and $q(y\mid x)$ defined over the space of images, we fit two distributions, $\mathcal{N}(\mu^{(p)}_{y\mid x}, \Sigma^{(p)}_{y\mid x})$ and $\mathcal{N}(\mu^{(q)}_{y\mid x}, \Sigma^{(q)}_{y\mid x})$ respectively, over featurizations of the image space. Note that these distributions are fitted separately for each $x_i$  in a test collection $\{x_i\}_{i=1}^N$, giving a collection of distributions $\{\mathcal{N}(\mu^{(q)}_{y|x_i}, \Sigma^{(q)}_{y|x_i})\}_{i=1}^N$. Finally, the objective is estimated as
\begin{equation}
    \mathbb{E}_{x\sim p(x)}\left[\mathbb{V}_{y\sim q(y\mid x, \Pi)}[y]\right]
    \approx \frac{1}{N} \sum_{i=1}^N \text{Tr}\left(\Sigma^{(q)}_{y|x_i}\right).
\end{equation}

% Reviewer 2: It is not clear to me that sampling diversity should be the metric used to validate the method. The science downstream of PSF deconvolution will most likely require uncertainty propagation, and thus the validation metric should be able to capture if the sampling method is calibrated....For calibration, you might want to look at Lemos et al. (2023), https://arxiv.org/abs/2302.03026

\section{Experiments}
\label{experiments}
\begin{figure*}
  \includegraphics[width=\textwidth]{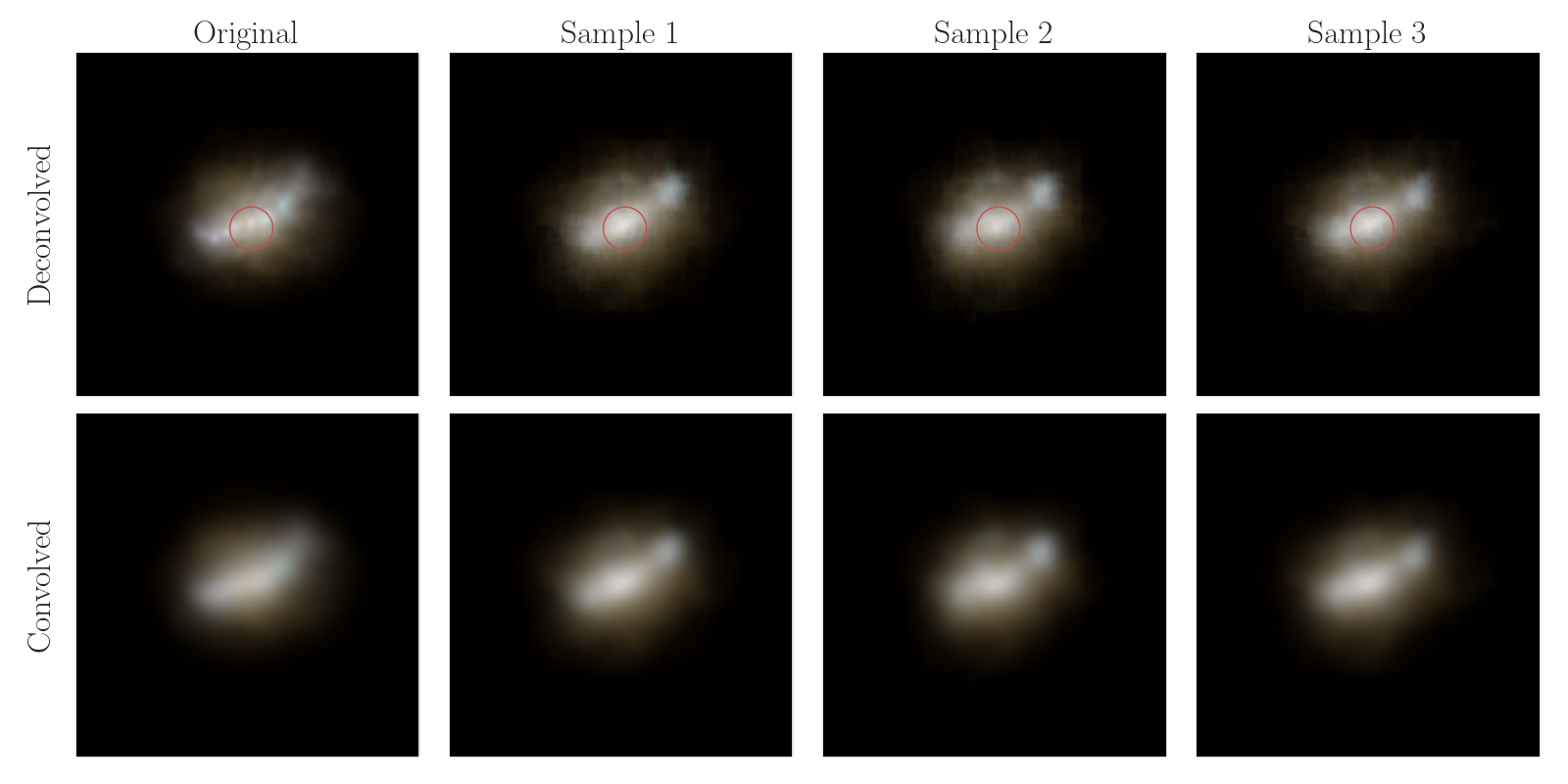}
  \caption{Candidate deconvolutions from the CVAE model. It can be seen that the generated samples shown are roughly valid inverse mappings by comparing them following convolution with the PSF (bottom row, columns 2--4) to the original PSF-convolved image (bottom row, column 1). However, high-frequency details, such as the belt of stars along the middle and other regions highlighted with red circles, are lost in reconstruction.}\label{fig:cvae_results}
\end{figure*}

\begin{figure*}
  \includegraphics[width=\textwidth]{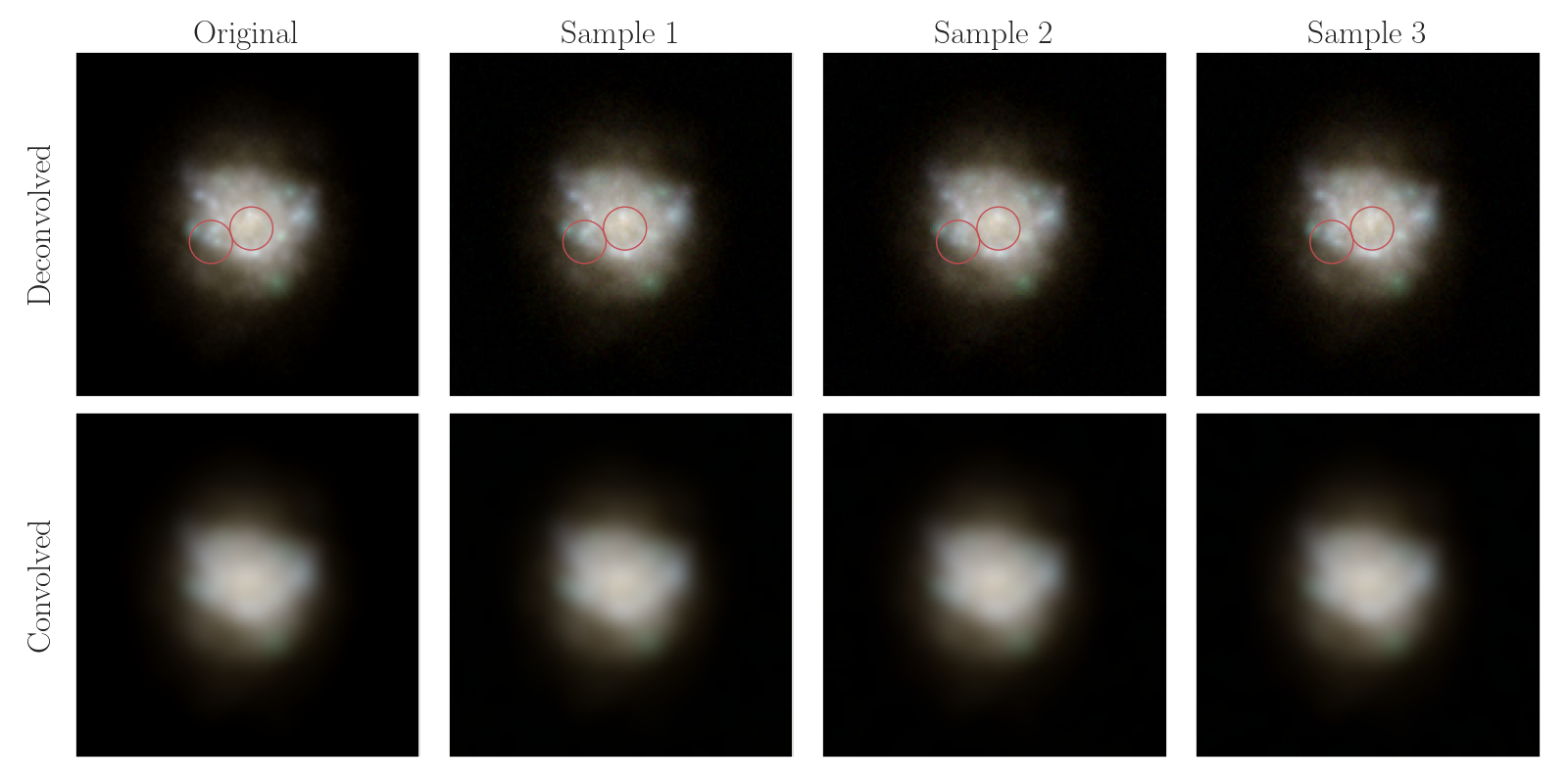}
  \caption{Candidate deconvolutions from the conditional diffusion model. It can be seen that the generated samples shown are roughly valid inverse mappings by comparing them following convolution with the PSF (bottom row, columns 2--4) to the original PSF-convolved image (bottom row, column 1). Additionally, high-frequency details (highlighted with red circles) are prominently captured in deconvolutions, enabling a greater diversity of reconstructions.}\label{fig:diff_results}
\end{figure*}

\begin{figure*}
  \includegraphics[width=\textwidth]{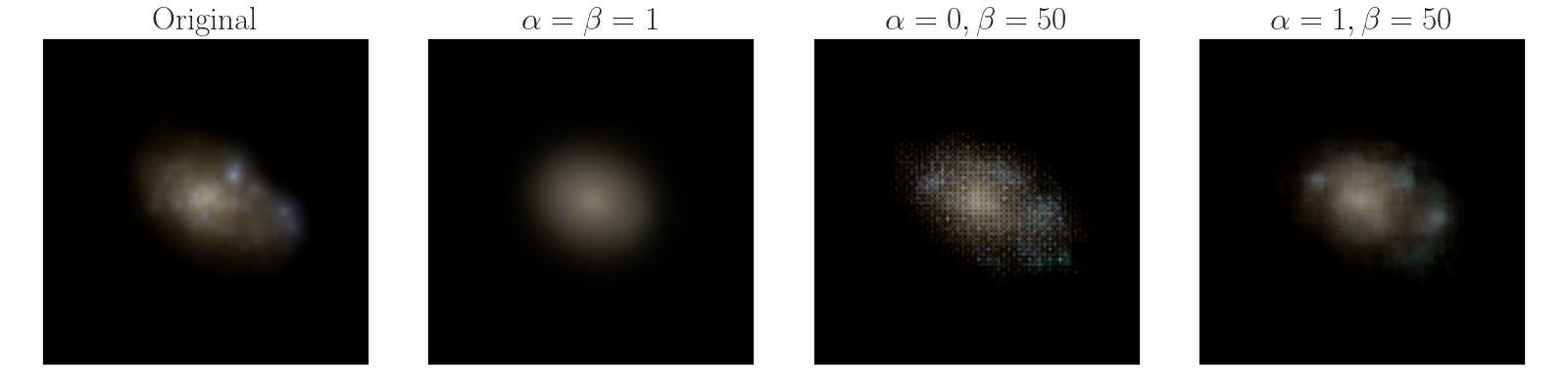}
  \caption{A representative sample from CVAEs trained to convergence with different choices of weights for the deconvolved and convolved image reconstructions, i.e. $\alpha$ and $\beta$ respectively. Observe that increasing the weight of the clean reconstruction tends to result in the loss of high-frequency detail.}\label{fig:training_weights}
\end{figure*}

We experiment with galaxy images produced by the IllustrisTNG simulator \cite{pillepich2018simulating}. This dataset provides a synthetic testbed similar in structure to the paired dataset of ground- and space-based telescope images that motivates our work. We construct a dataset consisting of tuples $\{(x_i, y_i, \Pi_i)\}_{i=1}^n$ by convolving each clean image $y_i$ with a PSF $\Pi_i$ sampled from a collection. We view the use of the clean image $y_i$ as an idealized surrogate for space-based telescopes.
% Reviewer 2: Noiseless debluring is not a realistic simulation of actual measurements of galaxy images on which this method would be applied. Noise is inherent in telescope measurements, and should be included in the simulation in some form or another. This work would greatly benefit from an application of the methods on real data. Furthermore, since the network is conditioned on the observation, it might suffer from domain shifts once used on real data. What are the thoughts of the authors on this, and how could this be mitigated?

Our dataset consists of 9718 images, each $128\times 128$ pixels, with 7774 used for training and 1944 reserved for validation. Evaluation of the aforementioned FID-like and variance metrics was performed on the validation set. Note that inference does not use the clean images. PSFs were taken to be discretized two-dimensional isotropic Gaussians on grids of size $10 \times 10$ with varying choices of $\sigma\in[2.0, 4.0]$ discretized in intervals of 0.5.
% Reviewer 1: I'm mildly worried about the limited variation amongst the PSFs in the data. They're all circular Gaussians, and the widths are discretized. It doesn't seem to be relevant for the comparison shown in this paper, but it might raise questions how well the DDPM can learn the deconvolution task for realistic ranges of PSF variations. Using samples from a Kolmogorov PSF would with some seeing distribution would allay such concerns.

All experiments were implemented in PyTorch \cite{paszke2019pytorch}. A standard U-Net architecture was used for the DDPM denoiser $\boldsymbol{\epsilon}_\theta(\cdot)$. Our implementation of the DDPM was based on the ``Conditional Diffusion MNIST'' project~\cite{pearce2023}.
Our CVAE employed a standard CNN-based architecture for both the encoder and decoder, with transposed convolutional layers used in the decoder. The observed PSF was included as an additional channel after it was encoded by a one-layer CNN network for both the DDPM and CVAE. The diffusion model was trained for 500 epochs with a minibatch size of 96, whereas the VAE model was trained for 600 epochs with a minibatch size of 128. For optimization, we used Adam \cite{kingma2014adam} with a learning rate of $10^{-4}$. We trained the diffusion model using one Nvidia A100 40G GPU, while the VAE model was trained using one Nvidia 2080 Ti GPU.
We used $T=950$ DDPM time steps. DDPM inference required 10 seconds per sample, while CVAE inference required just 0.01 seconds per sample.

Our CVAE was trained with asymmetric weights for deconvolved and convolved reconstructions, whose selection is justified by the results of Figure \ref{fig:training_weights}. To then assess the quality of the results according to Equation \ref{eqn:obj}, we first confirmed the validity of the samples across both the CVAE and the diffusion model by convolving them with the known PSF to ensure approximate recovery of the original images with a slack of $\epsilon=10^{-4}$. Although both accurately capture the low-frequency details, the CVAE fails to capture the high-frequency variation, resulting in visibly distinct reconstructed images compared to the originals (Figures~\ref{fig:cvae_results} and \ref{fig:diff_results}). 

Next, we discard samples with insufficient similarity between the reconstruction they imply and the original image. The proportion of samples retained for each method, which serves as a performance metric, is given in Table~\ref{table:metrics}.
We find that the diffusion model produces greater variety than the CVAE, as can be seen in Figures \ref{fig:cvae_results} and \ref{fig:diff_results} and in Table~\ref{table:metrics}. This variety manifests itself in subtle variations of high-frequency details that are equivalent under the forward convolution map. 

\begin{table}
\centering
\caption{Percent of retained samples and conditional variance metrics for samples generated by the diffusion and CVAE models. Quantitative results confirm the greater diversity apparent in visualizing the valid samples produced by the diffusion model over those from the CVAE.}
\label{table:metrics}
\begin{tabular}[t]{lll}
\toprule
Metric &             Diffusion &             CVAE \\
\midrule
% FID &  90.06 &  133.14 \\
Percent Retained & 100\%  & 53.9\%  \\
Variance & 15.86  & 15.13  \\
\bottomrule
\end{tabular}
\end{table}

\section{Discussion}
\label{discussion}
We investigated the sampling diversity of both CVAEs and conditional diffusion models that have been trained to perform PSF deconvolution. Diffusion models produce a greater diversity of valid deconvolution candidates compared to CVAEs, suggesting that they are preferable for downstream inference tasks.
In future work, we may apply conditional diffusion models to settings in which both the source PSF and the target PSF vary, by conditioning on the target PSF too. 
This extension would let us train high-fidelity disentangled galaxy models solely with images from ground-based surveys.
\bibliography{references}
\bibliographystyle{icml2023}

\end{document}